\documentclass[12pt,preprint]{aastex}

\def\be{\begin{equation}}
\def\ee{\end{equation}}

\def\ergs{{\rm\,erg\,s^{-1}}}
\def\msun{M_{\odot}}

\def\ergs{\rm \,erg\,s^{-1}}
\def\be{\begin{equation}}
\def\ee{\end{equation}}
\catcode`\@=11 
\def\@versim#1#2{\vcenter{\offinterlineskip
        \ialign{$\m@th#1\hfil##\hfil$\crcr#2\crcr\sim\crcr } }}
\def\lsim{\mathrel{\mathpalette\@versim<}}
\def\gsim{\mathrel{\mathpalette\@versim>}}

\shorttitle{Accretion-Jet Model}
\shortauthors{Feng Yuan, Wei Cui \& Ramesh Narayan}
\begin{document}

\title{An Accretion-Jet Model for Black Hole Binaries: 
Interpreting the Spectral and
Timing Features of XTE J1118+480}

\author{Feng Yuan}
\affil{Department of Physics, Purdue University, West Lafayette, IN 47907}
\email{fyuan@physics.purdue.edu}

\author{Wei Cui}
\affil{Department of Physics, Purdue University, West Lafayette, IN 47907}
\email{cui@physics.purdue.edu}

\author{Ramesh Narayan}
\affil{Harvard-Smithsonian Center for Astrophysics, 60 Garden Street,
Cambridge, MA 02138}
\email{narayan@cfa.harvard.edu}

\begin{abstract}

Multi-wavelength observations of the black hole X-ray binary XTE
J1118+480 have offered abundant spectral and timing information about
the source, and have thus provided serious challenges to theoretical
models.  We propose a coupled accretion-jet model to interpret the
observations. We model the accretion flow as an outer standard thin
accretion disk truncated at a transition radius by an inner hot
accretion flow.  The accretion flow accounts for
the observed UV and X-ray emission, but it substantially
under-predicts the radio and infrared fluxes, even after we allow for
nonthermal electrons in the hot flow.  We attribute the latter
components to a jet. We model the jet emission by means
of the internal shock scenario which is widely employed for gamma-ray
bursts. In our accretion-jet model of XTE J1118+480, the jet dominates
the radio and infrared emission, the thin disk dominates the UV
emission, and the hot flow produces most of the X-ray emission.  The
optical emission has contributions from all three components: jet,
thin disk, and hot flow. The model qualitatively accounts for
timing features, such as the intriguing positive and negative time
lags between the optical and X-ray emission, and the
wavelength-dependent variability amplitude.

\end{abstract}

\keywords{accretion, accretion disks --- black hole physics --- 
ISM: jets and outflows --- stars: 
individual (XTE J1118 + 480) --- X-rays: stars}

\section{Introduction}

Strong evidence now exists for black hole primaries in 15 X-ray novae
(also known as soft X-ray transients; McClintock \& Remillard
2004). One such source---XTE J1118+480---was
discovered with the All-Sky Monitor aboard the {\em Rossi X-Ray Timing
Explorer} (RXTE) on 2000 March 29 (Remillard et al.  2000). Subsequent
optical observations led to a measurement of the mass function,
$f(M)=6.00\pm0.36 \msun$, which represents a lower limit on the mass
of the compact primary and thus makes the source a secure black hole
candidate (BHC; McClintock et al. 2001a; Wagner et al. 2001).  XTE
J1118+480 is one of the best observed BHCs. It lies at an unusually
high Galactic latitude ($+62^{\circ}$), close to the ``Lockman Hole''
region.  The foreground absorption is extremely low (with $N_H \sim
0.7-1.3\times 10^{20} {\rm cm^{-2}}$; Hynes et al. 2000; McClintock et
al. 2001b), which allowed the detection of the source by the EUVE
satellite (Hynes et al. 2000). Simultaneous (or near-simultaneous)
observations were conducted, on multiple occasions, at radio,
infrared, optical, UV, EUV, and X-ray wavelengths, with
state-of-the-art instruments (Hynes et al. 2000; McClintock et
al. 2001b; Frontera et al. 2001; Chaty et al. 2003; McClintock et
al. 2003).

For clarity, we briefly summarize the main observational results here.
These include two aspects---spectral and timing features. The most
complete spectral energy distribution (SED) of XTE J1118+480 is shown
in Figures 1 and 2. The radio data are from
Fender et al. (2001) and the infrared to X-ray data from McClintock et
al. (2001) (all the data are associated with ``epoch 2'', when the
best simultaneous coverages were achieved; see Chaty et al. 2003 for a
summary of all observations).  The radio spectrum is well described by
a power-law of the form $F_{\nu}\propto \nu^{0.5}$.  Such a spectrum
is often thought to be typical of jet emission, although no jet has
been directly imaged, down to a limit of $<65$ $D ({\rm kpc}) ~{\rm
AU}$ (Fender et al. 2001), where $D$ is the distance to the
source. Note that we do not include in Figures 1 and 2 an
observational data point at 350 GHz (Fender et al. 2001), because
this measurement
was not done simultaneously with the others.  From IR to
UV, the spectrum is flat, with the {\em HST} spectrum exhibiting emission
lines.  Also, a Balmer jump is seen in absorption at $\nu\approx
10^{14.9}$ Hz (Hynes et al. 2000), implying that thermal emission
contributes substantially to the optical/UV band. The derived EUV
spectrum depends sensitively on the assumed $N_H$, which is still not
well constrained but probably lies in the range $N_H=1.0-1.3\times
10^{20}{\rm cm}^{-2}$ (McClintock et al.  2001b, 2004).
We take this uncertainty into account by requiring the model to stay
within the allowed range at EUV energies.  McClintock et al. (2001b)
fitted the X-ray spectrum
with a broken power-law. Above $\sim
2$ keV they obtained a photon index of $\approx 1.78$, 
but below $\sim 2$ keV the
spectrum appeared to be relatively harder.  However, calibration issues were
subsequently noted for the ACIS detectors used in the Chandra
observations\footnote{
see http://cxc.harvard.edu/cal/Acis/Cal\_prods/qeDeg/index.html.}.
This makes the spectrum uncertain at low energies. There 
is, in fact, independent evidence that the break at 2 keV may not be real.
XTE J1118+480 was observed many times with {\em BeppoSAX}, but
the X-ray spectra show no apparent deviation from a single power-law
at low energies (Frontera et al. 2001).

The main timing features include the following.  1) A quasi-periodic
oscillation (QPO) feature was detected in the X-ray light curve,
initially at a frequency $\nu\sim 0.08 $Hz (Revnitsev,
Sunyaev, \& Borozdin 2000), and was subsequently found to evolve (Wood
et al. 2000). The QPO was also detected in the optical
and UV bands at similar frequencies (Haswell et al. 2000; Yamaoka,
Ueda \& Dotani 2000). The fractional rms amplitude of the QPO is
$8-10$\% in the X-ray but only about 1\% at UV wavelengths (Hynes et
al. 2003, hereafter H03).  The fact that the same QPO frequency is
seen at optical, UV, and X-ray wavelengths indicates a common origin.
2) XTE J1118+480 also shows rapid aperiodic
variability at most wavelengths.  The variability amplitude is quite
large both in the X-ray and IR bands but is small in the optical/UV
band.  3) Correlation between emission at different wavelengths is
apparent (H03). In particular, cross-correlation analysis has revealed
some puzzling details in the correlation between the optical and X-ray
emission (Kanbach et al. 2001; H03; Malzac et al. 2003). In general,
the optical photons appear to lag the X-ray photons by $1-2$ s (see
H03, though with caveats). The lags are wavelength dependent; on
average a longer delay is seen at longer wavelengths.  On the other
hand, the cross-correlation function (CCF) also shows a ``precognition
dip'', i.e., the optical emission decreases about $2-5$ seconds {\em
before} the corresponding X-ray increase (Kanbach et al. 2001). At UV
wavelengths the ``dip'' appears to be weaker and the lag becomes
shorter, $\sim 0.5$ s (H03).  These complicated positive and negative
time lags between optical/UV and X-ray emission are not easy to
understand.  What is quite clear from the derived autocorrelation
functions (ACFs) is that the optical/UV emission is {\em not}
consistent with being due to the re-processing of X-ray photons by the
accretion disk, as is often assumed, because the ACF at optical/UV
wavelengths is {\em narrower} than that in X-rays (Kanbach et
al. 2001; Spruit \& Kanbach 2002; H03).

Several models have been proposed to explain the observed spectral and
temporal properties of XTE J1118+480.  Esin et al. (2001, hereafter
E01) explain the spectrum with an advection-dominated accretion flow
(ADAF) model, based on the work of Narayan (1996) and Esin,
McClintock, \& Narayan (1997).  They assume that the gas lost from the
secondary initially forms a standard thin disk outside a transition
radius $r_{\rm tr}$.  At $r_{\rm tr}$, the cool disk is truncated and
makes a transition to a hot accretion flow, described as an
ADAF (Narayan \& Yi 1994, 1995b; Narayan, Mahadevan \& Quataert 1998).
E01 satisfactorily explain the X-ray, EUV and UV spectra of the
source, but their model slightly under-predicts the optical flux and
significantly under-predicts the IR fluxes. They do not include radio
measurements in their work, but it is quite clear that their model
cannot account for the emission at radio wavelengths.

In contrast, Markoff, Falcke, \& Fender (2001) propose that the SED
of XTE J1118+480 is dominated
by synchrotron radiation from a jet, although they also need a truncated
accretion disk to explain the UV and EUV spectra.  Inside the truncation
radius, they assume that the accretion flow becomes an ADAF-like
accretion flow.  However, unlike E01, they ignore the radiation from
the ADAF.

No attempts have been made to explain the observed timing properties
with either of the above models. Merloni, Di Matteo \& Fabian (2000)
consider both spectral and timing data in their work, 
but their magnetic flare
model predicts that the disk emission should peak at
about $0.2$ keV, which is in disagreement with the {\em EUVE} and {\em
Chandra} data. Also, the model implies almost no time lag between
optical and X-ray photons, which seems to be at odds with the
measurements.  Recently, Malzac, Merloni \& Fabian (2004) have
proposed a time dependent, coupled disk-jet model for XTE J1118+480,
which has some resemblance to the model we discuss in this paper.
Whereas our model attempts to fit the spectral data
(see the following sections), Malzac et al. concentrate on understanding the
timing features. As pointed out by them, due to the complexity
of the time evolution of the accretion-jet system, detailed modeling
is impossible. They thus adopt a phenomenological approach. They model
the variability by assuming random fluctuations of the output power
from the disk and the jet, with the power being injected from a
reservoir of stored magnetic field.  By carefully choosing their
parameters, they are able to reproduce almost all the
observed timing features. These parameters can, in principle,
constrain the dynamics and geometry of the accretion flow.  One of their
interesting results is that they can rule out models in which
the energy budget is completely dominated by either the jet or
the accretion flow; rather, they favor a model in which both
components contribute.

In the present paper, we describe a coupled accretion-jet model 
to simultaneously account for both the spectral and timing properties of
XTE J1118+480.  We propose that the X-ray spectrum is produced mainly by
the ADAF-like hot accretion flow, whereas the radiation at longer
wavelengths comes from a jet (as in AGN). A similar idea has been
suggested previously (e.g., Hynes et al.  2000; McClintock et al. 2001;
Chaty et al. 2003).  In \S~2, we describe the model and discuss how it
can explain the SED of XTE J1118+480. In \S~3, we show that the
observed temporal properties can also be accommodated qualitatively
within the model.  We conclude in \S 4 with a summary and
discussion. We present in the Appendix technical details on
calculating the jet emission.

\section{Fitting the Spectrum}

\subsection{Accretion flow}

The accretion component of our model is implemented in nearly the same
manner as in E01, i.e., the accretion flow consists of an inner ADAF
and an outer thin disk. However, we have taken into account advances
in our understanding of the ADAF during the past ten years. First,
both numerical simulations (Stone, Pringle, \& Begelman 1999; Hawley
\& Balbus 2002; Igumenshchev et al. 2003) and analytical work (Narayan
\& Yi 1994, 1995a; Blandford \& Begelman 1999; Narayan et al. 2000;
Quataert \& Gruzinov 2000) indicate that probably only a fraction of
the gas that is available at large radius actually accretes
onto the black hole. The rest of the gas is either ejected from the
flow or is prevented from being accreted by convective motions.  The
details are likely to depend on the accretion rate.

We note that the outflow (and convection) is ultimately the result of
the accreting gas acquiring a positive Bernoulli parameter, as
emphasized by Narayan \& Yi (1994, 1995a).  Further, the effect is
strongest when the accretion rate is much below the threshold
above which ADAF ceases to exist.
Thus, accretion flows in highly under-luminous sources, like Sgr A* or
quiescent X-ray binaries, are expected to have strong
outflows.  On the other hand, the Bernoulli parameter decreases with
increasing radiative efficiency, and in fact becomes negative when the
radiative efficiency is large enough.  Therefore, for more luminous
systems like XTE J1118+480 in outburst and other X-ray binaries in the
low/hard state, which have relatively high accretion rates and
radiate fairly efficiently, we expect outflows and convection to be
less well-developed.  In the present paper, we allow for this effect
by adopting the following phenomenological prescription for the change
in mass accretion rate as a function of radius. We assume that, in the
hot flow, \be \frac{d{\rm ln}\dot{M}(r)}{d{\rm ln}r} \equiv s(r), \ee
where
\begin{mathletters}
\be s(r) = s_0f(r), \hspace{1cm} ({\rm if}~~  0\le f(r) \le 1), \ee
\be s(r) = 0, \hspace{2cm} ({\rm if}~~ f(r) \le 0). \ee  
\end{mathletters}
Here $s_0$ is a constant, which we set to $s_0=0.3$, as suggested by
our previous modeling of the highly advection-dominated source Sgr A*
(Yuan, Quataert \& Narayan 2003). The parameter $f(r)$ is the
advection factor of the accretion flow, defined as \be f(r) \equiv
\frac{q_{\rm adv}}{q_{\rm vis}}\equiv \frac{q_{\rm vis}-q_{\rm ie}}
{q_{\rm vis}}\,, \ee where $q_{\rm adv}, q_{\rm vis}$ and $q_{\rm ie}$
are the rates of energy advection, viscous heating, and Coulomb
collision cooling for the ions, respectively.  When the accretion rate
is very low, as in the case of Sgr A*, $q_{\rm vis} \gg q_{\rm ie}$,
so $f(r)=1$ and $s(r)=s_0$. In this case, from eq. (1) we have the
usual form, $\dot{M} = \dot{M}_0(r/r_{\rm tr})^{s_0}$, where
$\dot{M}_0$ is the accretion rate at the transition radius $r_{\rm
tr}$ (or the outer boundary of the ADAF). We adopt $s_0=0.3$, as in
the case of Sgr A* because the physics of the outflow should be the
same as long as $f(r)=1$ even though the accretion rates (in Eddington
units) can be quite different.  We should note, however, that our
results are not sensitive to the exact value of $s_0$.

A negative value of $f$ in eq. (2b) means that advection plays a
heating rather than a cooling role. In this case, the hot accretion
flow is described by a luminous hot accretion flow (hereafter LHAF)
model, which is a natural extension of an ADAF to higher accretion
rates (Yuan 2001, 2003). From ADAF to LHAF, both $\dot{M}$ and the
radiative efficiency increase continuously and smoothly. Yuan \&
Zdziarski (2004) argue that for luminous X-ray sources, such as the
low/hard states of some BHCs and Seyfert 1 galaxies, the
luminosity may be above the highest luminosity an ADAF can reach but
could be accommodated by an LHAF. We allow for an LHAF in this work,
because it is unclear at present which solution, ADAF or LHAF, applies
to XTE J1118+480. We simply refer to both the ADAF and LHAF solutions
as {\em hot accretion flows}.

We calculate the global solution of the hot accretion flow, starting
at $r_{\rm tr}$ and integrating inward. The numerical details may be
found in Yuan (2001).  One main difference with E01 is that we solve
the radiation hydrodynamics equations self-consistently, and thus we
obtain the exact value of $f(r)$ at each radius.  In contrast, E01
used the approximation that $f(r)$ has a constant average value at all
radii.  On the other hand, we treat Comptonization within a 
local approximation, whereas E01 computed the Comptonization globally
using the method described in Narayan, Barret \& McClintock (1997).
The radiation processes we consider include bremsstrahlung,
synchrotron emission, and the Comptonization of both synchrotron
photons from the hot accretion flow and soft photons from the cool
disk outside $r_{\rm tr}$. The emission from the outer cool disk is
modeled as a multicolor blackbody spectrum. The effective temperature
as a function of radius is determined by the viscous dissipation and
the irradiation of the disk by the inner hot flow.

Yuan \& Zdziarski (2004) found that to explain the X-ray emission of
most black hole X-ray binaries, $\alpha\ga 0.1$ is required (see also
Narayan 1996).  We fix $\alpha$ and the magnetic parameter $\beta$
(defined as the ratio of the gas pressure to the sum of gas and
magnetic pressure) at their ``typical'' values: $\alpha=0.3$,
$\beta=0.9$. We set $\delta=0.5$, i.e., $50\%$ of the viscous dissipation
heats electrons directly. The exact value of $\delta$ does not
affect our results very much since the required $\dot M$ to model 
XTE J1118+480 in outburst
is relatively high, so the main heating mechanism
for electrons is energy transfer from ions via Coulomb collisions.
In this sense $\alpha$, $\beta$ and $\delta$ are not free
parameters, though we should emphasize that large uncertainties exist
here.  We set the mass of the black hole
at $M=8 \msun$, the distance to the source at $D=1.8$ kpc, and the
binary inclination $\theta=70^{\circ}$ (McClintock et al. 2001a;
Wagner et al. 2001).  Following E01, we estimate 
the outer radius of the cool disk
using Paczy\'{n}ski's formula (Paczy\'nski 1971): $r_{\rm
out}=3\times 10^4r_s(10\msun/M)^{2/3}$, where $r_s\equiv2GM/c^2$ is
the Schwarzschild radius of the black hole. The free parameters of the
accretion flow are the transition radius $r_{\rm tr}$, the accretion
rate at the transition radius $\dot{M}_0$, and an outer boundary
condition ---the temperature of the accretion flow at $r_{\rm tr}$
(Yuan 1999).

Figure 1 shows the spectral fitting results obtained with the
accretion flow model.  The values of the parameters are:
$\dot{M}_0=0.05\dot{M}_{\rm Edd}$, $r_{\rm tr}=300 r_s$.  The X-ray
emission is produced by Comptonization in the hot flow. The main seed
photons are from synchrotron emission by the thermal electrons in the
hot flow (as assumed in the original ADAF model of Narayan \& Yi 1995b),
as opposed to the blackbody emission of the thin disk.
This is also consistent with the prediction of Wardzinski \& Zdziarski (2000)
given that XTE J1118+480 is not very luminous. For more luminous
sources, the seed photons may be dominated by blackbody emission from
the thin disk. The EUV and UV in the model are mostly 
from the outer thin disk.  The fit is
satisfactory, although the optical fluxes are slightly
under-predicted. The fact that the UV/optical emission is dominated by
the thin disk explains the presence of Balmer jump absorption and
emission lines and reprocessing features in the data (\S1). The IR
and radio fluxes are significantly under-predicted, however (ref. Fig. 2).
Figure 3 shows the profiles of the advection factor $f(r)$ and the
fractional mass accretion rate $\dot{M}(r)/\dot{M}_0$ as a function of
radii. We see that $f(r)$ is positive over much of the flow except
near $r_{\rm tr}$. Since most of the radiation comes from the inner
region where $f(r)>0$, the solution is in the ADAF rather than LHAF
regime, consistent with E01. This is because the luminosity of XTE
J1118+480 is not high.

While our results are in general agreement with those of E01, there
are two noteworthy differences.  First, our value of $r_{\rm tr}
(=300r_s)$ is significantly larger than that of E01 ($r_{\rm
tr}=55r_s$).  This discrepancy is mainly due to two reasons.  First,
E01 adopted a no-torque boundary condition at $r_{\rm tr}$ while we
apply this condition at the marginally stable orbit of the black hole.
Second, in E01 the mass accretion rate of the thin disk follows
$\dot{M}(r)=\dot{M}_0(1-r_{\rm tr}/r)$ while we simply use
$\dot{M}(r)= \dot{M}_0$. Both differences are related to the physics
of the transition of the accretion flow at $r_{\rm tr}$, which is
highly uncertain at present, so it is not clear which approach is more
appropriate. As a comparison, $r_{\rm tr}=352 r_s$ in Chaty et al.
(2003) who fitted the EUV spectrum, while $r_{\rm tr}=17 r_s$ in
Frontera et al. (2001; 2003) who fitted the iron line and reflection
features. The second difference between our model and E01 is that the
value of $\dot{M}$ in E01 ($\dot{M}_0=0.02\dot{M}_{\rm Edd}$) is
significantly smaller than ours ($\dot{M}_0=0.05\dot{M}_{\rm
Edd}$). This is primarily because (1) we include an outflow in our
calculations so that the accretion rate in the inner region is smaller than
that at $r_{\rm tr}$ (see Fig. 3 where $\dot M \sim 0.03 \dot M_{\rm
Edd}$ near the black hole in our model, close to E01's
value); and (2) we use the pseudo-Newtonian potential of Paczy\'nski \&
Wiita (1980), while E01 used the general relativistic solution of
Popham \& Gammie (1998) in calculating the radial velocity of the
accretion flow.  As shown by Narayan et al. (1998), the latter gives
higher luminosity for the same accretion rate.

To account for the under-prediction of the
IR and radio fluxes, we first consider the effect of nonthermal
electrons in the hot accretion flow.  Since the inflowing gas is
collisionless, processes such as MHD turbulence, reconnection, and
weak shocks can accelerate electrons and generate a nonthermal tail at
high energies in the electron distribution function. Yuan, Quataert \&
Narayan (2003) found that the radio spectrum of Sgr A*,
which was under-predicted by a pure ADAF model with only thermal
electrons, can be explained if roughly $1\%$ of the electron energy
is in nonthermal electrons.  We tested this idea for XTE J1118+480.
The dotted line in Figure 2 shows the (absorbed) synchrotron
emission from nonthermal electrons.
We see that there is a sharp cut-off below about $10^{13}$ Hz, so that
the emission from nonthermal electrons is unable to fit the radio and
IR fluxes.  This result is not sensitive to how much energy the
nonthermal electrons have. In the case of Sgr A*, the emission from
nonthermal electrons extends to much lower frequency and forms a
power-law spectrum. The difference between Sgr A* and XTE J1118+480 is
that in the latter case the density is several orders of magnitude
higher.  Therefore, the magnetic field in XTE J1118+480 
is much stronger and the lowest
frequency that the power-law electrons emit is much higher.
We conclude that the accretion flow alone cannot account for
the low-frequency spectrum of XTE J1118+480 at radio and IR
wavelengths.  Some other component, most likely a jet, is required.

\subsection{Coupled Accretion-Jet Model}

Jets are thought to occur in the low/hard state of BHCs (see Fender
2004 for a review).  There have been many papers on the emission of
radio jets in active galactic nuclei (e.g., Blandford \& K\"onigl
1979; Ghisellini, Maraschi, \& Treves 1985; Falcke 1996). In the
present paper, following Spada et al. (2001), we adopt the internal
shock scenario widely used in interpreting gamma-ray burst (GRB)
afterglows (e.g., Piran 1999). The details of the model of the jet
radiation are described in Appendix A. Briefly, we assume that, near
the black hole, a fraction of the accretion flow is transferred into
the vertical direction to form a jet.  Since the radial velocity of
the accretion flow near the black hole is supersonic, a standing shock
should occur at the bottom of the jet due to the bending. From the
shock jump conditions, we calculate the properties of the postshock
flow, such as the electron temperature $T_e$. We assume a constant 
$T_e$ in the jet, which is clearly over-simplified, since adiabatic
expansion will cause the electrons to cool.  However, the assumption
has very little effect on the results because the jet emission is
dominated by the nonthermal electrons discussed below. We assume that
the jet has a conical geometry with half opening angle $\phi$, and
that the bulk Lorenz factor of the jet $\Gamma_{\rm j}$ is
independent of distance from the black hole.  We further assume that
internal shocks occur due to the collision of shells with different
$\Gamma_{\rm j}$.  These shocks accelerate a fraction of the electrons
into a power-law energy distribution with index $p=2.24$ (e.g., Kirk
et al. 2000). The steady state energy distribution of the accelerated
electrons is carefully determined since it is important for
calculating the emitted spectrum. The effect of radiative cooling
is considered in this process. Following the widely adopted approach
in the study of GRBs, we specify 
the energy density of accelerated electrons and
amplified magnetic field by two free parameters,
$\epsilon_e$ and $\epsilon_B$.  We then calculate the radiative
transfer by both thermal and power-law electrons in the jet, although
we find that the latter plays a dominant role.  Only synchrotron
emission is considered since Compton scattering is not important in
this case (see also Markoff, Fender \& Falcke 2001).

The thin solid line in Figure 2 shows the emission of the jet. The
parameters are: mass loss rate in the jet $\dot{M}_{\rm jet}=2.5\times
10^{-4}\dot{M}_{\rm Edd}$, which is about $0.5\%$ of the accretion
rate in the accretion disk, $\phi=0.1$, $\epsilon_e=0.06$,
$\epsilon_B=0.02$, bulk Lorenz factor of the jet $\Gamma_{\rm
j}=1.2$, and length of the jet $\sim 13 $ AU.  The values of
$\epsilon_e$ and $\epsilon_B$ are well within the typical range
obtained in GRB afterglows (e.g., Panaitescu \& Kumar 2001; 2002), and
the length of the jet is consistent with the observed upper limit of
$65 D({\rm kpc})$ AU. The value of $\Gamma_{\rm j}$ is well within the
range obtained by combining observations and numerical simulations:
$\Gamma_{\rm j} \la 1.67$ (Gallo, Fender \& Pooley 2003).
We see from Figure 2 that the jet emission fits the low-frequency radiation
very well.  The IR flux is dominated by the jet, while from optical to
UV, the jet becomes less important. The
contribution of the jet to EUV and X-rays is negligible. We should
point out that the solution shown is not unique and that the jet
parameters are not as well constrained as those of the accretion
flow. However, the results are not very sensitive to the values of the
jet parameters.

It is interesting to check whether a pure thermal jet can also explain
the data. We find that we can get an equally good fit to the spectrum
if we adjust the geometry and $T_e (z)$ profile of the jet
carefully. In this model, we only need a tiny fraction of the gas in
the accretion flow, $\sim 0.003 \%$, to go into the jet. 
However, the required
temperature is very high, $T_e \sim 10^{10}$ K. In addition, the jet
velocity has to be very low, $\sim 100~{\rm km~s^{-1}}$; otherwise,
the required magnetic field in the jet becomes unrealistically
large. Such a low speed close to the black hole seems unphysical.

\section{Interpreting the Timing Features}

\subsection{QPOs}

Numerous models have been proposed to explain the QPO phenomenon
in X-ray binaries (see review by van der Klis 2000). In some models, 
the QPO frequency is associated with the Keplerian frequency of the
accretion flow at a special radius---the transition radius
$r_{\rm tr}$ in our case. For example, Giannios \& Spruit (2004; 
see also Rezzolla et al. 2003) suggest that the QPO can be excited
by the interaction of the inner hot accretion flow and outer thin
disk.  The QPOs then result from the basic p-mode oscillations of the inner
hot accretion flow, with frequency roughly equal to the Keplerian
frequency at $r_{\rm tr}$. The Keplerian frequency at $r_{\rm
tr}=300r_s$ is $\sim 0.22$ Hz, which is roughly consistent with the
observed QPO frequency of $\sim 0.1$ Hz. Because the entire
region of the hot flow oscillates collectively at the same frequency,
and the emission from the hot flow contributes somewhat at both
optical/UV and X-ray (see Fig. 1), the QPO should be observable at
both optical/UV and X-ray wavelengths with the same frequency.
Wood et al. (2000) find that the QPO frequency in XTE J1118+480
increases from 0.07 to 0.15 Hz during the
outburst, while the 2-10 keV X-ray flux slowly rises and then
decreases. Our calculations do not show such a non-monotonic relationship,
so the evolution of the QPO remains a puzzle. We should emphasize that
the non-monotonic change of the QPO frequency with the flux is
not universal among BHCs. In fact, for most sources, the
correlation seems to be monotonic (e.g., Cui et al. 1999).

\subsection{Variability amplitude}

The variability amplitude from the jet is expected to be large, both
from internal shocks and from possible instabilities in the jet. The
hot accretion flow is thermally marginally unstable, so any
perturbations in it will survive and move inward, as shown by
numerical simulations (Manmoto et al. 1996) and analytical work (Yuan
2003).  However, the growth timescale of the perturbations is longer than the
accretion timescale, so the hot accretion flow 
is not threatened by the instability.
The simulations further show that the simulated flux variation can
account for the observed substantial variability observed in BHCs.
On the other hand, the intrinsic variability
of emission from the thin disk should be very weak because the
characteristic timescale is many hours even at $r_{\rm tr}$, i.e.,
much longer than the observed $\sim$ seconds or minutes variability
timescale (e.g., Kanbach et al. 2001). The only source
of variability of the thin disk emission is due to the reprocessing of
the variable X-ray radiation, but the contribution of this component
is very weak.

With the above knowledge, we can qualitatively understand 
variability amplitudes at different wavelengths. Large variability in
the IR and X-ray bands is natural because the IR emission is dominated
by the jet and the X-ray emission by the hot flow.  As the emission
from the disk becomes more important in the optical and UV, the
source varies less in these bands. The correlation between optical/UV and
X-ray is easily understood because the hot accretion flow contributes
in both bands.  H03 find that the spectral energy distribution (SED)
of the variable component of the emission is roughly a power law,
which they argued as being consistent with optically thin synchrotron
radiation. However, given the fact that the rms amplitudes were derived 
from light curves with the same time resolution, it is actually
not straightforward to interpret the result, since the intrinsic
variability timescale at different wavelengths should be quite
different. Moreover, the physical origin of the variability is likely
to be complicated (e.g., Malzac, Merloni, \& Fabian 2004). We note
that a power-law SED of the variability does not arise naturally
in a pure jet model (e.g., Markoff, Falcke \& Fender 1999). For
instance, if we assume that the variability is caused by fluctuations
in $\dot{M}_{\rm jet}$, such a model would predict a power-law index
of $0.8$, which is the same as the X-ray spectral index, while the
measured index of the variability spectrum is $\sim 0.59$ (H03).

\subsection{Correlations between optical/UV and X-ray}

Suppose there is a perturbation due to an instantaneous increase of
$\dot{M}_0$. The X-ray flux will increase.  The increase in $\dot M$
will propagate inward with the accretion flow, and eventually 
will lead to an
increase in the mass loss rate and thus the optical/UV emission from
the jet.  This could explain why the optical/UV variability {\em lags}
the X-ray variability. Quantitatively, we find that in our model the
optical/UV emission from the jet 
comes mainly from regions at a distance
of about $d \sim 6000r_s$ from the black hole. This corresponds to a
propagation time of $\sim d/c\sim 1.2$ s, consistent with the measured
$\sim 1-2$ s lag. The size of the optical emission region is $\sim
2d\phi\sim 1200 r_s$, where $\phi$ is the half opening angle of the
jet. The corresponding light crossing time is $1200 r_s/c\approx
0.1$s, consistent with the shortest variability timescale $\sim 100 $
ms seen in the optical (e.g., Kanbach et al. 2001).  Since the
emission at longer wavelengths originates from regions farther away,
the time lag should increase with increasing wavelength.

As for the negative lag, we note that, for the parameters of our model
(Fig. 1), an increase of $\dot{M}$ in the hot accretion flow results
in a {\em decrease} of the optical/UV flux, as shown in Figure 4. 
The optical/UV emission from the hot accretion
flow is mainly due to self-absorbed synchrotron emission, which
depends on the profiles of $T_e$ and optical depth $\tau$. For our
model, an increase in $\dot{M}$ causes a decrease in the flux.
In our model, the optical emission comes from $\sim 35r_s$, the UV
from $\la 10 r_s$, and the X-rays from $\sim 7-8 r_s$. So when
$\dot{M}_0$ increases, the optical flux will first decrease, then the
UV will decrease, and finally the X-ray flux will increase. This might
be the origin of the negative lag of the optical/UV, as well as the
negative correlation, and may also explain why the lag in the UV is
shorter than in the optical. Since the emission from the hot accretion
flow contributes less at shorter wavelengths in the
optical/UV regime (see Fig. 1), we can also understand why the dip becomes weaker
at shorter wavelengths.  Since the IR flux from the hot accretion flow
does not vary with varying $\dot{M}_0$ (see Fig. 4), we predict that
such a negative lag should be absent between IR and X-ray.

Quantitatively, however, we are not able to account for the magnitude
of the negative lags. The viscous timescale at $\sim 35r_s$ is $\sim
0.1$ s, which is more than 20 times smaller than the observed $2-5$ s
negative lag seen in the optical. This might be due to an approximation
in the outer boundary condition we assume for the global solution.
For technical reasons, we set
the angular velocity of the flow at $r_{\rm tr}$
to be substantially
sub-Keplerian, $\Omega (r_{\rm tr}) \sim 0.5 \Omega_{\rm k}$, even
though it should be super-Keplerian (Abramowicz, Igumenshchev, \&
Lasota 1998); otherwise, the viscous dissipation would be negative.
and the solution would be unphysical (see also Manmoto, Mineshige, \&
Kusunose 1997). Since the centrifugal force is the dominant factor
determining the radial velocity of the accretion flow, our
approximation makes the radial velocity much larger than it should
actually be and thus lead to a shorter viscous timescale. In addition,
the viscosity parameter $\alpha$ may be smaller than the value we
adopted, which will again result in a longer viscous timescale.

Finally, we note that an increase of $\dot{M}$ in the cool thin disk
will obviously result in an increase in the optical/UV emission.
However, such an increase is unlikely to be seen in the
cross-correlation analysis, since the accretion timescale in the thin
disk is on the order of hours.

\section{Summary and Discussion}

The observational data on XTE J1118+480 is almost unique among all
current BHCs. The spectral and timing information impose very strong
constraints on theoretical models and provide us with an
opportunity to understand in detail the inflow/outflow processes around black
holes. In this paper we explain how these observations can be
understood in the context of a coupled accretion-jet model. In our
model, the accretion flow is described as a geometrically thin cool
disk outside a transition radius $r_{\rm tr}$ and a
geometrically-thick hot accretion flow inside $r_{\rm tr}$, as in the
model of E01.  We adopt a phenomenological prescription for the magnitude of
the mass outflow from the hot accretion flow (eqs. 1--3).
The free parameters
describing the accretion flow are the transition radius $r_{\rm tr}$,
the mass accretion rate at $r_{\rm tr}$, $\dot{M}_0$, and the outer
boundary condition at $r_{\rm tr}$. The spectrum due to the accretion
flow alone is shown in Figure 1. The X-ray emission is dominated by
Comptonization of synchrotron photons in the hot accretion flow, and
both the EUV and UV are dominated by the cool disk.  The fit is quite
satisfactory in these bands. The optical flux is slightly
under-predicted, however, and the IR and radio spectra are
significantly under-predicted (Fig. 2). These results are very similar
to those of E01.

Obviously, we require an additional component in the model to 
explain the IR and radio fluxes.
We first consider the possibility of nonthermal
electrons in the hot accretion flow, but find that this idea does not
work. We stress, however, that the failure does not mean that there
are no non-thermal electrons in hot accretion flows.
Such electrons might, for instance, be responsible for
the ``hard tail'' in the spectrum of Cyg X-1 in the low/hard state
(McConnell et al. 2000). 

Having eliminated non-thermal electrons as an explanation for the low
frequency emission of XTE J1118+480, we argue that the radiation must
originate in a jet. Assuming that a small fraction of the mass in
the accretion flow is transferred to the jet, we calculate the jet
emission using the internal shock scenario that is widely adopted in
the study of GRB afterglows.  The results of the accretion-jet
model are shown in Figure 2.  We find that the radiation from the jet
can account for all of the radio and IR emission and part of the
optical/UV emission.  The required mass loss rate in the jets is about
0.5\% of the accreted matter.

The coupled accretion-jet model not only explains the spectrum, it
also qualitatively explains many of the timing features observed in
XTE J1118+480.  These features include the frequency of QPO; the
similarity of the QPO frequency in optical/UV/X-ray bands (\S3.1); the
dependence of the variability amplitude on wavelength (\S3.2); and the
positive and negative time lags between optical/UV and X-ray (\S3.3).
Quantitatively, however, we are not able to account for the magnitude
of the negative time lag between X-ray and optical/UV
(\S 3.3).

It is interesting to examine the energetics of the accretion flow and
the jet in our model. The total accretion power is $P_{\rm
acc}=\dot{M}_0c^2\sim 5 \times 10^{38}\ergs$ and the power lost in the
outflow is $P_{\rm outflow}\equiv P_{\rm acc}-\dot{M}(r_s)c^2=3.6
\times 10^{38}\ergs$.  The X-ray luminosity emitted by the hot
accretion flow is $L_{\rm x-ray} \sim 2\times 10^{36}\ergs$, the
optical/UV luminosity emitted by the thin disk is $\sim 2\times
10^{36}\ergs$, and the jet power is $P_{\rm jet} =\Gamma_{\rm j}^2
\dot{M}_{\rm jet}c^2\sim 3.6 \times 10^{36}\ergs$, which is $\sim 2$
times $L_{\rm x-ray}$. For comparison, Malzac et al.  (2004) require
$P_{\rm jet}/L_{\rm x-ray} \sim 10$ to reproduce the main timing
features of XTE J1118+480, while Fender et al. (2001) estimate $P_{\rm
jet}/L_{\rm x-ray} \ga 0.2$. The luminosity emitted by the jet in 
our model is $L_{\rm jet}\sim    
2\times 10^{35}\ergs$, so the radiative efficiency of the jet is $\sim
0.055$, roughly consistent with the estimate of $\sim 0.05$ by Fender
et al. (2001) but larger than the value of $\sim 0.003$ in Malzac et
al. (2004). So there are differences in both the 
value of $P_{\rm jet}/L_{\rm x-ray}$ and the efficiency of the jet
between our model and that of Malzac et al. (2004). One reason for 
the discrepancy is that Malzac et al. {\em assume} the optical flux
to be completely dominated by synchrotron emission from
the jet, while our detailed modeling shows that the contribution from
the accretion flow and the jet are comparable in the optical band
(Fig. 2).  Thus, more power from the jet is required in their model.
In addition, the estimated value of
$L_{\rm jet}$ in Malzac et al. (2004) is only $ 5 \times 10^{34}\ergs$, 
which is $\sim$ 4 times smaller than ours. This is because they integrate the jet 
emission from radio to optical, while in our model, the jet emission extends 
up to X-rays (ref. Fig. 2). 

Assuming $P_{\rm acc}-P_{\rm outflow}=1.4 \times
10^{38}\ergs$ to be the accretion power in the inner region of the
accretion flow from which most of the X-ray and jet power originate,
we see that only $L_{\rm x-ray}/(P_{\rm acc}-P_{\rm outflow})
\sim 1\%$ is released through the X-ray emission and
$\sim 2\%$ channeled into the jet, while most of the accretion power
is stored in the accretion flow and advected into the black hole. In
other words, XTE J1118+480 is radiatively quite inefficient, in
agreement with the conclusion of Malzac et al. (2004).  The small
ratio of the jet power to the accretion power also justifies our
approximation that the jet has very little effect on the global
solution of the hot accretion flow. We should point out that some
uncertainties exist in the above estimations concerning the jet since
the jet parameters in our model are not as well constrained as the
parameters of the accretion flow.

Several other caveats also need to be mentioned.  First, we adopt a
pseudo-Newtonian potential rather than the exact general relativistic
approach when we calculate the dynamics of the hot accretion
flow. Secondly, we adopt a sub-Keplerian angular velocity at the
transition radius whereas the rotation here should be super-Keplerian.
The main effect of these two approximations is that the radial
velocity in the hot flow is larger than it should actually be, and
thus the density is smaller than the ``correct value''. We believe
that most of the effect is absorbed in the accretion rate parameter
$\dot{M}_0$. But the approximations do affect some quantitative result
such as the time lag between optical/UV and X-ray.  Thirdly, we have
not explored fully the parameter space. The values of several
parameters such as $\alpha, \beta,$ and $\delta$ are fixed in our
calculations (to 0.3, 0.9 and 0.5, respectively).  Investigating their
effects in detail by surveying their entire parameter space would be
very time-consuming and is beyond the scope of the paper.

The philosophy of this paper is that the hard X-ray emission comes
from the hot accretion flow via thermal Comptonization, and that the
contribution from the jet is negligible in this band.  This is
different from the model of Markoff et al. (2001) in which synchrotron
radiation from the jet dominates in X-rays.  We note that many details
of the X-ray observations of BHCs have been successfully explained
with a hot accretion flow model (see the review by Zdziarski \&
Gierli\'{n}ski 2004) and it remains an open question whether the jet
model can do equally well.  Poutanen \& Zdziarski (2002) and Zdziarski
et al. (2003) have pointed out some difficulties with the jet
proposal. For example, the non-thermal synchrotron emission in this
model cannot produce a sharp enough cut off at high energies, and the
predicted spectrum is not as hard as the spectra observed in many
BHCs. Also, the jet model should yield X-ray variability virtually
independent of energy, which is in strong disagreement with the
observational data.  Finally, it is unclear if the
model can explain the various timing features of XTE J1118+480
described in this paper.

Of course for some black hole sources, the emission from the jet dominates over
the accretion flow in the X-ray band. BL Lacs are a well-known class
of objects where this situation is known to exist. In previous work we
have discussed this possibility also for two other sources, Sgr A* and
NGC~4258 (Yuan, Markoff, \& Falcke 2002; Yuan et al. 2002).
In the case of NGC~4258, the jet emission dominates the accretion flow
because we require a significant fraction of the accretion flow to be
transfered to the jet, $\dot{M}_{\rm jet}/ \dot{M}_0 \approx 10-25
\%$, which is more than $\sim 20$ times higher than in XTE J1118+480.
Such a high value perhaps
implies that the black hole in NGC~4258 is very rapidly spinning.  In
the case of Sgr A*, the value of $\dot{M}_{\rm jet}/\dot{M}_0$ is
similar to XTE J1118+480, but the X-ray emission from the jet is
comparable to the accretion flow. This is because the accretion rate
(in Eddington units) in Sgr A* is much lower.  The flux from the
accretion flow, which comes from (multi-order scattering) Comptonization
radiation, increases much faster with the accretion rate than that
from the jet, which is from synchrotron and (one-order scattering)
synchrotron-self-Compton emission.  Therefore, the ratio of jet to
disk flux increases with decreasing Eddington-scaled accretion rate.

Recently a very interesting correlation between radio and X-ray fluxes
has been discovered in GX 339-4.  The correlation 
extends over more than three
decades in X-ray flux (Corbel et al. 2003). Such a correlation likely
exists in other BHCs and even in AGN (Gallo, Fender, \& Pooley 2003;
Merloni, Heinz \& Di Matteo 2003; Falcke, K\"ording, \& Markoff
2004). The correlation is sometimes used as evidence for a jet origin
for the X-ray emission of BHCs, e.g., Markoff et al. (2003). However,
Heinz (2004; see also Merloni, Heinz \& Di Matteo 2003) recently 
pointed out that if the electron energy 
spectrum is not too steep and if  radiative losses are included, both
of which are required by observations, the jet model cannot explain the 
radio---X-ray correlation. Merloni, Heinz \& Di Matteo (2003) further showed
that the X-ray emission is unlikely to be produced by radiatively efficient
accretion (as in the sandwiched corona+disk geometry); 
rather, the accretion flow
must be radiatively inefficient. Our preliminary investigations
indicate that the radio---X-ray correlation can be explained in the context of
our accretion-jet model (Yuan \& Cui 2004, in preparation).

\begin{acknowledgements}
We thank Drs. S. Chaty, R.I. Hynes, and J.E. McClintock for 
providing us with the data. Helpful comments by Drs. S. Chaty and J.E. 
McClintock are acknowledged. This work was supported in
part by NASA grants NAG5-9998 and
NAG5-10780 and NSF grant AST-0307433.

\end{acknowledgements}

\vfill\eject
\begin{appendix}
\section{The Internal Shock Model for Jet Radiation}

We adopt the internal shock scenario to calculate the emission from the 
jet, similar to Spada et al. (2001). We are interested only in the 
time-averaged spectrum. 
Following Blandford \& K\"onigl (1979), we assume the jet is in 
conical geometry, with semi angle of $\phi$ whose axis makes an 
angle $\theta$ with the direction of observer. The jet has a constant
velocity, characterized by a bulk Lorenz factor of $\Gamma_{\rm j}$,
and has constant plasma temperature. The mass loss rate in 
the jet is,
\be
\dot{M}_{\rm jet}=\pi z^2\phi^2 \rho(z)v_{\rm j}
\ee
The quantity  $\rho(z)$ is the mass density of the jet plasma at distance $z$
from the black hole, measured in the jet-comoving frame.  

The main assumption in the internal shock scenario is that the central power
engine produces energy which is channelled into jets in an intermittent way,
thus faster shells will catch up with slower ones and internal shocks 
are formed in the jet. The minimum distance the shells propagate before
collision occurs is $z_0\sim \Gamma_{\rm j}^2 r_s$ (Piran 1999; Spada et al. 2001).
Our results are not sensitive to its exact value. 

The bulk Lorenz factor of steady jets in BHCs  
is likely only mildly relativistic (Fender 2004), e.g., 
$\Gamma_{\rm j} \la 1.67$ from Gallo, Fender \& Pooley 2003. 
In this case, for an adiabatic index of $4/3$, 
the energy density of the internal shock is (Piran 1999),
\be
e_2=\gamma_2 n_2 m_pc^2
\ee
where $\gamma_2=\sqrt{(\Gamma_{\rm j}^2+1)/2}$
 is the Lorenz factor of the formed internal shock,
$n_2=(4\gamma_2+3)n_{\rm 1}$ is the post-shock number density with $n_{\rm 1}$
is the preshock number density in the jet determined by eq. (A1). 

The shock will heat plasma in the jet, generate/amplify the magnetic field, and
accelerate a small fraction of electrons into relativistic energy. We assume that 
the fraction of accelerated electrons in the shock is $\xi_e$ and fix $\xi_e=1\%$. 
Given the uncertainty in shock physics, as the usual approach, we
introduce two dimensionless parameters, $\epsilon_e$ and $\epsilon_B$,
which measure the fraction of the comoving internal energy
of the internal shock stored in the accelerated electrons and magnetic field.
Obviously, $\xi_e$ and $\epsilon_e$ are not independent. 

Assume that the {\em injected} electrons after the shock
acceleration have a power-law distribution with 
index $p$,
\be
n_{\rm pl}(\gamma_e)d\gamma_e=N_{\rm pl}(p-1)\gamma_e^{-p}, \hspace{1cm}
\gamma_{\rm min}\le\gamma_e\le\gamma_{\rm max}
\ee
We set $p=2.24$, according to the results of relativistic shock
acceleration of Bednarz \& Ostrowski (1998) and
Kirk et al. (2000). In this case ($p>2$), we have
\be
N_{\rm pl}=\xi_en_2\gamma_{\rm min}^{p-1}.
\ee
Now we calculate the value of $\gamma_{\rm min}$. We have,
\be
N_{\rm pl}m_ec^2\frac{p-1}{p-2}\gamma_{\rm min}^{2-p}=\epsilon_e
U_{\rm sh}=\epsilon_e(\gamma_2-1)n_2m_pc^2
\ee
where $U_{\rm sh}=(\gamma_2-1)n_2m_pc^2$ is the internal 
energy density of the internal shock. From the above 
equation and the definition of $\xi_e$, we can obtain
\be
\gamma_{\rm min}=(\gamma_2-1)\frac{p-2}{p-1}\frac{m_p}{m_e}\frac{\epsilon_e}{\xi_e}
\ee
The value of $\gamma_{\rm max}$
is not important if we are not interested in the fitting
the X-ray spectrum of XTE J1118+480 with jet emission.
When radiative cooling of relativistic electrons is important, as
in the present case of XTE J1118+480, the {\em steady} distribution of 
electrons is different from eq. (A3). Defining a ``cooling
Lorenz factor'' $\gamma_c$ at which the radiative timescale $t_{\rm rad}$
is equal to 
the dynamical timescale $t_{\rm dyn}$ at distance $z$ in the jet,
\be
t_{\rm rad}=\frac{3}{4}\frac{8\pi m_e c}{\sigma_T \gamma_c\beta_e^2B^2}
=t_{\rm dyn}=\frac{z}{c},
\ee
then depending on the
relative value of  $\gamma_{\rm min}$ and $\gamma_c$, there will be two cases
for the steady distribution.
When $\gamma_{\rm min} > \gamma_c$, we have,
\begin{mathletters}
\be n_{\rm pl}(\gamma_e)d\gamma_e=N_{\rm pl}(p-1)\gamma_c\gamma_{\rm min}^{1-p}
\gamma_e^{-2} d\gamma_e, \hspace{1cm}  
\gamma_c\le\gamma_e \le \gamma_{\rm min}, \ee
\be n_{\rm pl}(\gamma_e)d\gamma_e=N_{\rm pl}(p-1)\gamma_c\gamma_e^{-(p+1)}
d\gamma_e, \hspace{2cm} 
\gamma_e \ge \gamma_{\rm min}. \ee
\end{mathletters}
When $\gamma_{\rm min} < \gamma_c$, we have,
\begin{mathletters}
\be n_{\rm pl}(\gamma_e)d\gamma_e=N_{\rm pl}(p-1)\gamma_e^{-p}d\gamma_e, 
\hspace{1cm}  
\gamma_{\rm min}\le\gamma_e \le \gamma_{\rm c}, \ee
\be n_{\rm pl}(\gamma_e)d\gamma_e=N_{\rm pl}(p-1)\gamma_c\gamma_e^{-(p+1)}d\gamma_e,
\hspace{2cm} 
\gamma_e \ge \gamma_{\rm c}. \ee
\end{mathletters}

The magnetic field generated/amplified by the shock is determined by,
\be
\frac{B^2}{8\pi}=\epsilon_BU_{\rm sh}=\epsilon_B(\gamma_2-1)n_2m_pc^2
\ee

Since most of electrons may still be in thermal distribution, we need
to consider their role in emitting and absorbing photons. To this
purpose, we need to know their temperature. One constraint comes
from the following consideration. If the jet is formed 
at the innermost region of the accretion flow, within the sonic
point at $\sim 10r_s$, since the accretion flow is supersonic, when it is 
bended into the vertical direction to form the jet, 
a standing shock should occur. Note that the global solution of ADAF 
(e.g., Narayan, Kato \& Honma 1997) does not find shocks.
Our assumption of the bending shock is not in conflict 
with this result since jet was not considered in that calculation. On the 
other hand, shock is found in the general relativistic MHD numerical
simulations of jet formation (e.g., Koide et al. 2000).
>From the global solution of the accretion flow,
we know the values of preshock quantities. Applying the shock jump 
conditions at the jet radius, we then be able to calculate the postshock
quantities, including the electron temperature (see Yuan, Markoff, \& Falcke
2002 for details). Adiabatic expansion will cause the electrons
to cool while the internal shocks 
in the jet will further heat the electrons. But for simplicity,
we do not consider these effects, since we find the radiation
from the power-law electrons dominate over that from thermal ones.

Now we are ready to calculate the emission from the jet. The emissivity
from each location in the jet is,
\be
I_{\nu}^{\rm out}(z)=\int^{\tau_0}_0e^{-\tau}S_{\nu}(\tau)d\tau
\approx \frac{j_{\rm th}+j_{\rm pl}}{\alpha_{\rm th}+\alpha_{\rm pl}}
\left(1-e^{-\tau_0}\right)
\ee
where $\tau$ is the optical depth along the line of sight in the jet, 
$S_{\nu}=(j_{\rm th}+j_{\rm pl})/(\alpha_{\rm th}+\alpha_{\rm pl})$ 
is the source function, including the emission and absorption from both
thermal ($j_{\rm th}, \alpha_{\rm th}$)
and power-law ($j_{\rm pl}, \alpha_{\rm pl}$) electrons
in the jet. We then integrate the
emission from different distance in the jet to obtain the total emission.
The relativistic effects is taken into account in the calculation.
There is a remaining important point when we do the integration, that is, 
we should not integrate all of the volume of the jet. A 
``volume filling factor'' $f_{\rm sh}(<1)$ should be introduced. The value of
$f_{\rm sh}$ is very uncertain. It obviously depends on the ``spatial
density'' of the internal shocks in the jet. In addition, 
the generated/amplified magnetic field in the shock may survive
for only a short time, this will further decease its value.
We set $f_{\rm sh}=0.1$ in our model. Fortunately this value is not very
important since it can be absorbed in $\dot{M}_{\rm jet}$.

\end{appendix}

{} 

\clearpage

\begin{figure}
\epsscale{0.90}
\plotone{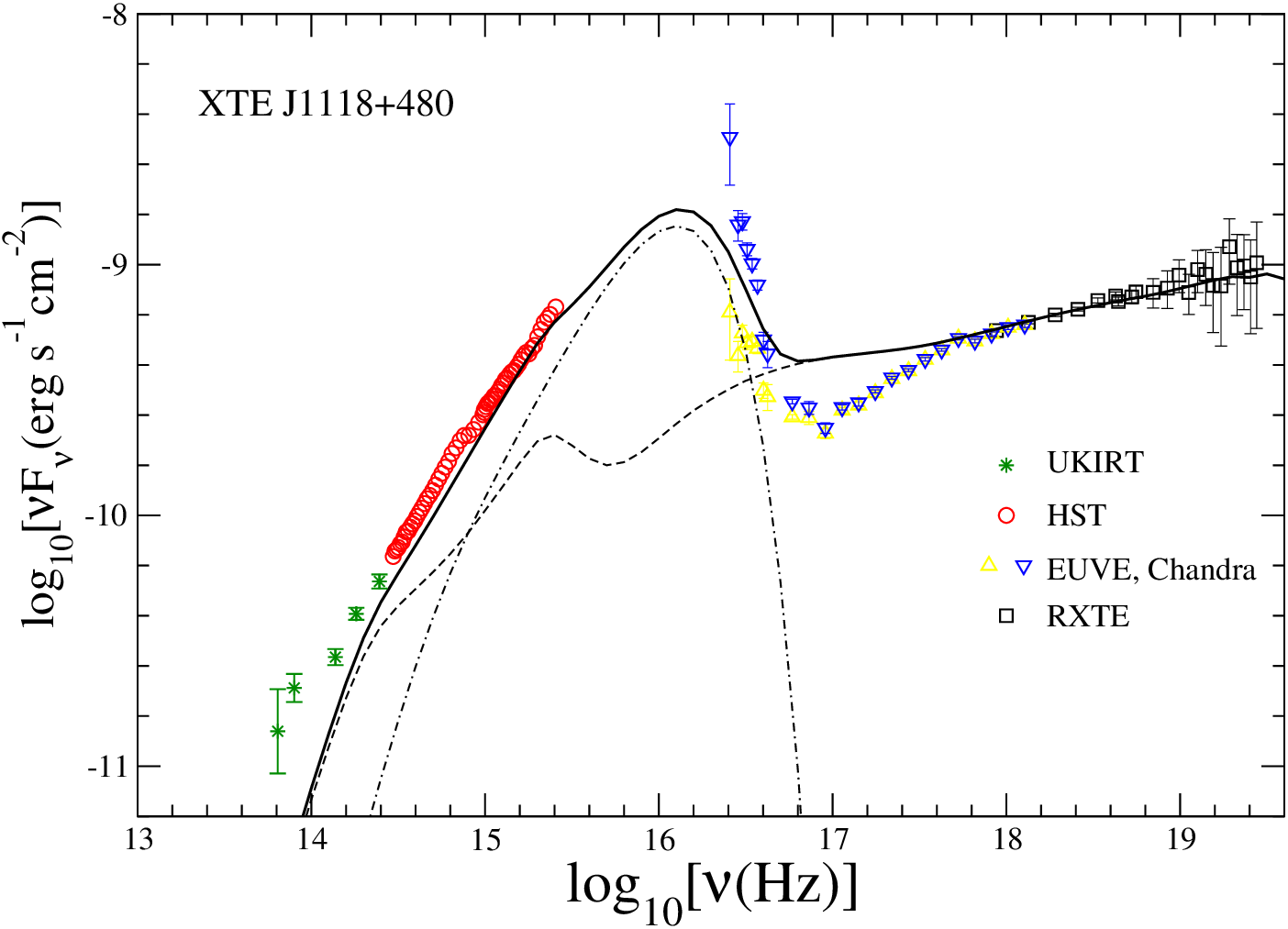}
\vspace{.0in}
\caption{Spectral modeling results for XTE J1118+480. The fit was made
with a model consisting of an inner hot accretion flow and an outer
cool thin disk. The parameters of the model are $r_{\rm tr}=300 r_s$,
$\dot{M}_0=0.05\dot{M}_{\rm Edd}$, $\alpha=0.3, \beta=0.9,
\delta=0.5$. The dashed line shows the emission from the inner hot
accretion flow, the dot-dashed line shows the emission from the outer
cool disk, and the solid line shows the sum of the two. The model
explains the EUV and X-ray data quite well, slightly under-predicts
the optical/UV, and significantly under-predicts the IR and radio
fluxes (the radio data are shown in Fig. 2). Note that two sets of EUV
data are shown, for two different choices of $N_H$. The X-ray spectral
break at $\sim 10^{17.7}$ Hz may not be real (see text for details).}
\end{figure}
\clearpage
\begin{figure}
\epsscale{0.90}
\plotone{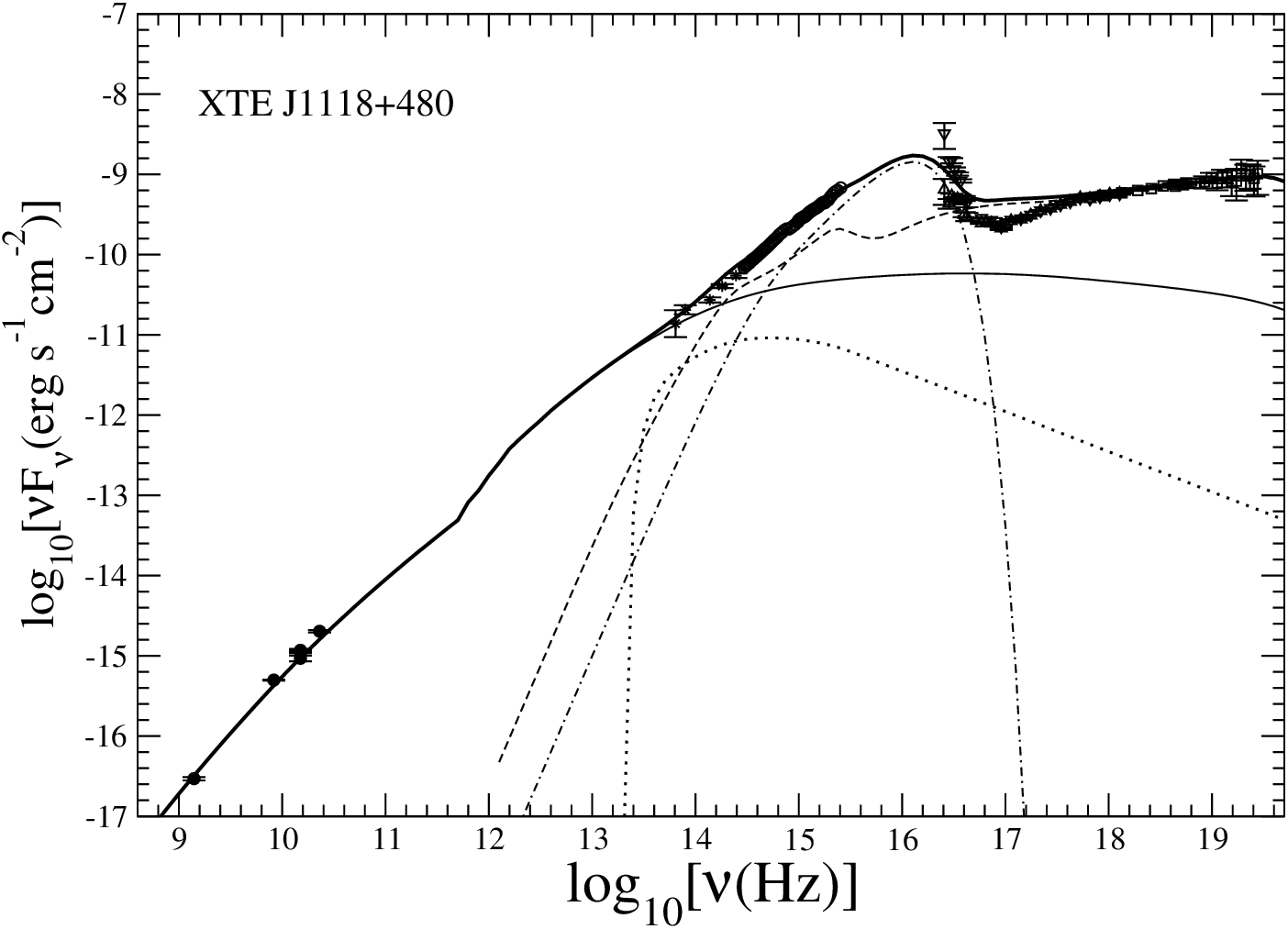}
\vspace{.3in}
\caption{Accretion-jet model of XTE J1118+480. The dashed and
dot-dashed lines show the emission from the hot and cool accretion
flows, respectively, as in Fig. 1. The thin solid line shows the
emission from the jet.  The sum of the three components, shown by the
thick solid line, explains the spectrum all the way from radio to
X-rays.  The dotted line shows the synchrotron emission from power-law
electrons that might be present in the hot accretion flow.}
\end{figure}
\clearpage
\begin{figure}
\epsscale{0.90}
\plotone{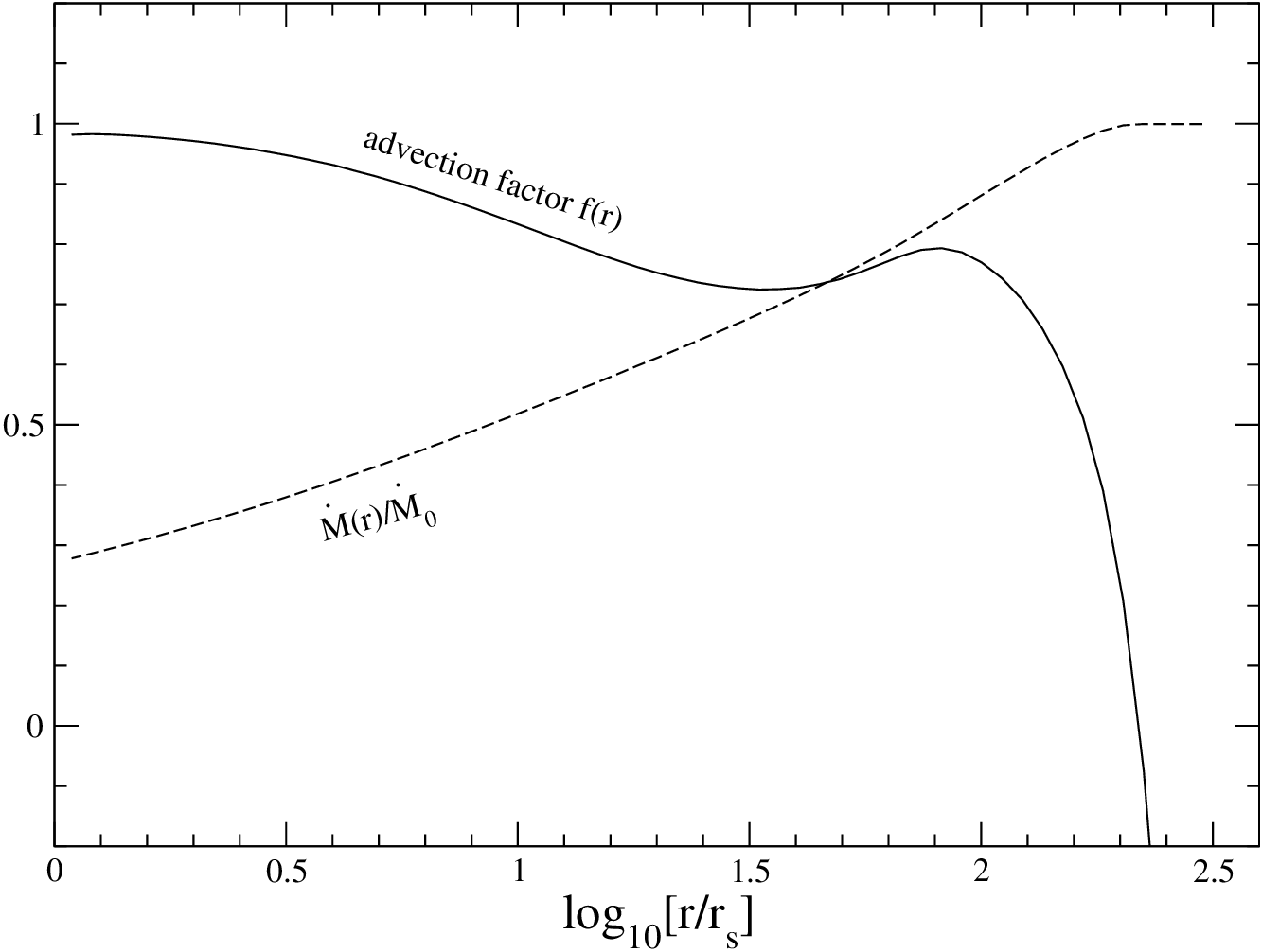}
\vspace{.2in}
\caption{Advection factor $f$ (defined in eq. 3) and the scaled mass
accretion rate, $\dot{M}(r)/\dot{M}_0$, as a function of radius for
the hot accretion flow model shown in Fig. 1. Negative values of $f$
indicate that the accretion flow is in the ``LHAF'' regime rather than
the ADAF regime at these radii. The solution is basically an ADAF.}
\end{figure}
\clearpage
\begin{figure}
\epsscale{0.90}
\plotone{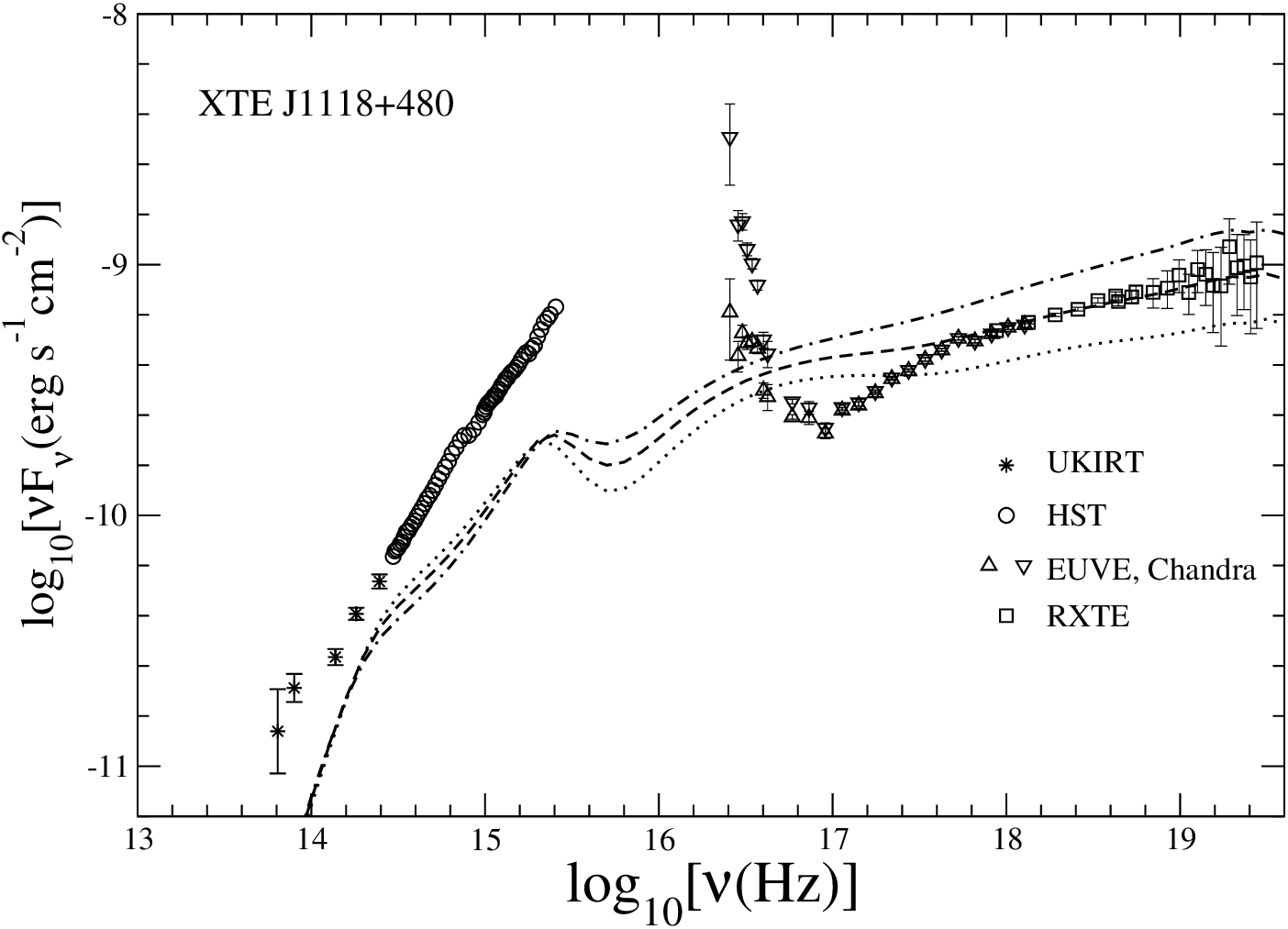}
\vspace{.2in}
\caption{Model spectra from the hot accretion flow for three choices
of $\dot{M}_0/\dot{M}_{\rm Edd}$: 0.04 (dotted), 0.05 (dashed), and
0.06 (dot-dashed).  All other parameters are held fixed.}
\end{figure}

\end{document}